\documentclass[%
 reprint,
 superscriptaddress,
 amsmath,amssymb,
 aps,
]{revtex4-2}
\usepackage{graphicx}
\usepackage{dcolumn}
\usepackage{bm}
\usepackage{lipsum}
\usepackage{cleveref}
\usepackage{amsmath,color}

\begin{document}


\title{Supervised perceptron learning vs unsupervised Hebbian unlearning:\\
approaching optimal memory retrieval in Hopfield-like networks
}

\author{Marco Benedetti}
 \thanks{These two authors contributed equally}
\affiliation{
 Dipartimento di Fisica, Sapienza Università di Roma, P.le A. Moro 2, 00185 Roma, Italy
 }

\author{Enrico Ventura}
\thanks{These two authors contributed equally}
\affiliation{
 Dipartimento di Fisica, Sapienza Università di Roma, P.le A. Moro 2, 00185 Roma, Italy
 }
 \affiliation{Laboratoire de Physique de l'Ecole Normale Sup\'erieure, ENS, Universit\'e PSL, CNRS, Sorbonne Universit\'e, Universit\'e de Paris, F-75005 Paris, France
}

\author{Enzo Marinari}
\thanks{Corresponding authors: enzo.marinari@uniroma1.it, giancarlo.ruocco@uniroma1.it, francesco.zamponi@ens.fr}
\affiliation{
 Dipartimento di Fisica, Sapienza Università di Roma, P.le A. Moro 2, 00185 Roma, Italy, and CNR-Nanotec and INFN Sezione di Roma
 }

\author{Giancarlo Ruocco}
\thanks{Corresponding authors: enzo.marinari@uniroma1.it, giancarlo.ruocco@uniroma1.it, francesco.zamponi@ens.fr}
\affiliation{
 Dipartimento di Fisica, Sapienza Università di Roma, P.le A. Moro 2, 00185 Roma, Italy
 }

\author{Francesco Zamponi}
\thanks{Corresponding authors: enzo.marinari@uniroma1.it, giancarlo.ruocco@uniroma1.it, francesco.zamponi@ens.fr}
\affiliation{Laboratoire de Physique de l'Ecole Normale Sup\'erieure, ENS, Universit\'e PSL, CNRS, Sorbonne Universit\'e, Universit\'e de Paris, F-75005 Paris, France
}


\date{\today}

\begin{abstract}
The Hebbian unlearning algorithm, i.e. an unsupervised local procedure used to improve the retrieval properties in Hopfield-like neural networks, is numerically compared to a supervised algorithm to train a linear symmetric perceptron. We analyze the stability of the stored memories: basins of attraction obtained by the Hebbian unlearning technique are found to be comparable in size to those obtained in the symmetric perceptron, while the two algorithms are found to converge in the same region of Gardner's space of interactions, having followed similar learning paths. A geometric interpretation of Hebbian unlearning is proposed to explain its optimal performances. 
Because the Hopfield model is also a prototypical model of disordered magnetic system, it might be possible to translate our results to other models of interest for memory storage in materials.
\end{abstract}

\maketitle 


\section{\label{sec:1}Introduction}

Hopfield-like neural networks are very successful models of associative memory~\cite{hop82,Amit}. In this framework, the network is composed of $N$ binary neurons $\{\sigma_i\}_{i=1}^N=\pm 1$, and it is used to store $P=\alpha N$ binary {\it memories} $\{\xi_i^\mu\}_{\mu=1}^P=\pm 1$ where the \textit{load} $\alpha=P/N$ will be used as a control parameter of the model. By \textit{memorize} we mean that the network must be able to reconstruct the memories on the basis of their noise-corrupted version. This is achieved by endowing it with an appropriate dynamics, such that fixed point attractors with finite basins of attractions are present in close proximity to the memories. If the network is initialized close enough to one of the stored memories, the dynamics will drive it to the corresponding attractor, reducing the number of misaligned spins. 
In general, the dynamics is given by a zero temperature asynchronous Monte Carlo dynamics~\cite{Amit,peretto}, in an energy landscape given by the Hamiltonian
\begin{equation}
    H[\sigma]=-\frac{1}{2}\sum_{i, j}J_{ij}\sigma_i \sigma_j \ .
    \label{eq:general_hamiltonian}
\end{equation}
At every step we pick randomly a site $i$ and update the spin according to
\begin{equation}
\label{eq:dynamics}
    \sigma_i\to\text{sign}\Big(\sum_{j(\neq i)}^NJ_{ij}\sigma_j\Big)\ .
\end{equation}
The details of the dynamics depend on how the synaptic interaction matrix $J$ is shaped. One of the most influential models in the field is that introduced by  Hopfield~\cite{hop82}, where the coupling matrix is built according to Hebb's prescription~\cite{Amit, Hebb}:
\begin{equation}
    J_{ij}=\frac{1}{N}\sum_{\mu=1}^P\xi_i^\mu\xi_j^\mu \,, \quad J_{ii}=0.
    \label{eq:hebbian_learning}
\end{equation}
The phase diagram of the model has been extensively studied, and includes a recognition phase for ${\alpha<\alpha_c\sim0.14}$, where a fraction of the fixed point attractors of the dynamics are very close to the memories~$\vec{\xi}^\mu$~\cite{AGS}. Proximity between two configurations $\vec{\sigma}^1$ and $\vec{\sigma}^2$ is naturally measured in terms of their overlap
\begin{equation}
    m=\frac{1}{N}\sum_{i=1}^N\sigma_i^1\sigma_i^2 \ .
\end{equation}
It is also well known that the disorder of the model implies the existence of dynamic multistability, i.e. a rough landscape of local minima of the energy having non-vanishing overlap with the memories~\cite{spur}. Such \textit{spurious states} are thus fixed points of the dynamics and they proliferate when $\alpha \geq \alpha_c$.
In the Hopfield model the overlap between  attractors and  memories is smaller than 1 for any extensive load $\alpha\neq 0$, and deteriorates as the memory load is increased, up to a discontinuous transition to zero when $\alpha_c$ is reached, which is often referred to as \textit{blackout catastrophe}~\cite{Amit,Marcooo}. Hence, the error correcting performance of a Hopfield network can never be perfect, and the memories are not themselves fixed point of the dynamics.

Many strategies have been proposed in order to extend the capacity of Hopfield-like models, and improve error correcting performance in the retrieval phase \cite{Marcooo,kanter,Dotsenko,PS,Hop83,Agliari, max}. In this note we highlight some surprising similarities between two popular options: the {\it supervised} symmetric perceptron algorithm (SP)~\cite{Gardner88,Gardner89,Forrest} and the {\it unsupervised} Hebbian unlearning (HU) algorithm~\cite{Hop83,VH90,VH92,VH94}. The paper is organized as follows. In Sec.~\ref{sec:2} we introduce the algorithms.  In Sec.~\ref{sec:3} and \ref{sec:4} we present our main results, and in Sec.~\ref{sec:5} we propose a key to interpret those results. Finally in Sec.~\ref{sec:6} we summarize our findings.

\section{\label{sec:2}The algorithms}
In this section we detail how the symmetric perceptron and Hebbian unlearning algorithms operate, and how they improve the performance of the Hopfield model.

We note first that the problem of storing the memories as fixed points of the dynamics is mathematically equivalent to finding a set of couplings that satisfy the constraints (recall that $J_{ii}=0$)
\begin{equation}
\label{eq:const}
    \xi^\mu_i=\text{sign}\Big(\sum_{j}^NJ_{ij}\xi^\mu_j\Big)\ , \qquad \forall \mu, i \ .
\end{equation}
Written in this way, this becomes a supervised learning problem, in which binary input vectors $(\vec{\xi}^\mu)_j:=\xi^\mu_j$ must be correctly associated to binary labels $\xi_i^\mu$ by a collection of $N$ single-layer perceptrons with weights given by $N\text{-dimensional}$ vectors $(\vec{J_i})_j:=J_{ij}$, $1\leq i,j\leq N$, i.e. $\xi_i^\mu = \text{sign}(\vec{J_i}\cdot \vec\xi^\mu)$.

An elegant way to solve this problem is to recast it in terms of a linear regression~\cite{minsky,Gardner88}.
For every memory and every spin, we define $N\text{-dimensional}$ \textit{patterns} 
\begin{equation}
    \vec{\eta}_i^\mu=\xi_i^\mu\vec{\xi}^\mu \ .
\end{equation}
By multiplying both sides of Eq.~\eqref{eq:const} by $\xi_i^\mu$, the constraints can be expressed as
\begin{equation}
\Delta_i^\mu=\frac{\vec{J_i} \cdot \vec{\eta}_i^\mu }{|\vec{J_i}|} \geq 0 \ .
\end{equation}
The quantities $\Delta^\mu_i$ are called \textit{stabilities}~\cite{krauth}.
A stronger version of the constraint, $\Delta^\mu_i>k$, is satisfied when the vector $\vec{\eta}_i^\mu$ lies on the positive side of the oriented plane normal to $\vec{J_i}$ and passing through the origin, at a distance greater than $k$ from it. In these terms, the problem of perfectly stabilizing a number $P$ of $N\text{-dimensional}$ memories has been factorized into $N$ independent linear separation problems, each classifying a number $P$ of $N\text{-dimensional}$ vectors. The parameter $k\geq 0$ can be used to tune the stabilities of the memories~\cite{Gardner88}.

\subsection{The symmetric perceptron}
\label{sec:2a}

The symmetric perceptron algorithm solves this separation problem under the condition $J_{ij}=J_{ji}$. 
It is defined by the following procedure for constructing the matrix $J$~\cite{Gardner89,Forrest}:
\begin{itemize}
    \item Initialize $J_{ij}$ to a symmetric matrix.
    \item Update $J_{ij}$ until convergence according to 
    \begin{equation}
    \begin{aligned}
        J_{ij}&\to J_{ij}+\lambda\sum_{\mu=1}^P(\epsilon^\mu_i+\epsilon^\mu_j)\xi^\mu_i\xi^\mu_j, \quad J_{ii}=0 \ , \\
        \epsilon^\mu_i&=\theta(-\Delta^\mu_i+k) \ .
    \end{aligned}
    \label{eq:gardner_update_rule}
    \end{equation}
\end{itemize}
Notice that symmetry in the coupling matrix is preserved by the algorithm. 
This algorithm is {\it supervised}, in the sense that it needs to be provided with the full set of memories $\{\vec{\xi}^\mu\}$ at every step.   
The \textit{masks} $\epsilon^\mu_i$ are defined in such a way that the algorithm stops when $\Delta^\mu_i>k$ for all $\mu$ and $i$, where the stability threshold $k$ is a parameter of the algorithm. $\lambda$ is the tunable {\it learning rate} of the algorithm. 
If $k$ exceeds a critical threshold $k_{max}(\alpha)$, no symmetric matrix $J$ exists satisfying the stability requirement $\Delta^\mu_i>k$, and the algorithm does not converge (unsatisfiable, or UNSAT, phase). On the other hand, when $k<k_{max}(\alpha)$ such coupling matrices exist, and the algorithm will converge to one of them in a finite number of steps (satisfiable, or SAT, phase). In the limit case $k=k_{max}(\alpha)$, only one coupling matrix meets the stability requirements, and the algorithm is supposed to converge to it, independently on the value of the learning rate $\lambda$ and of the initial $J$. It has been shown that, as one could expect, increasing $k$ towards $k_{max}$ leads not only to more stable memories, but also to larger basins of attraction for each memory~\cite{Forrest}.

\begin{figure}
\centering
\includegraphics[width=.4\textwidth]{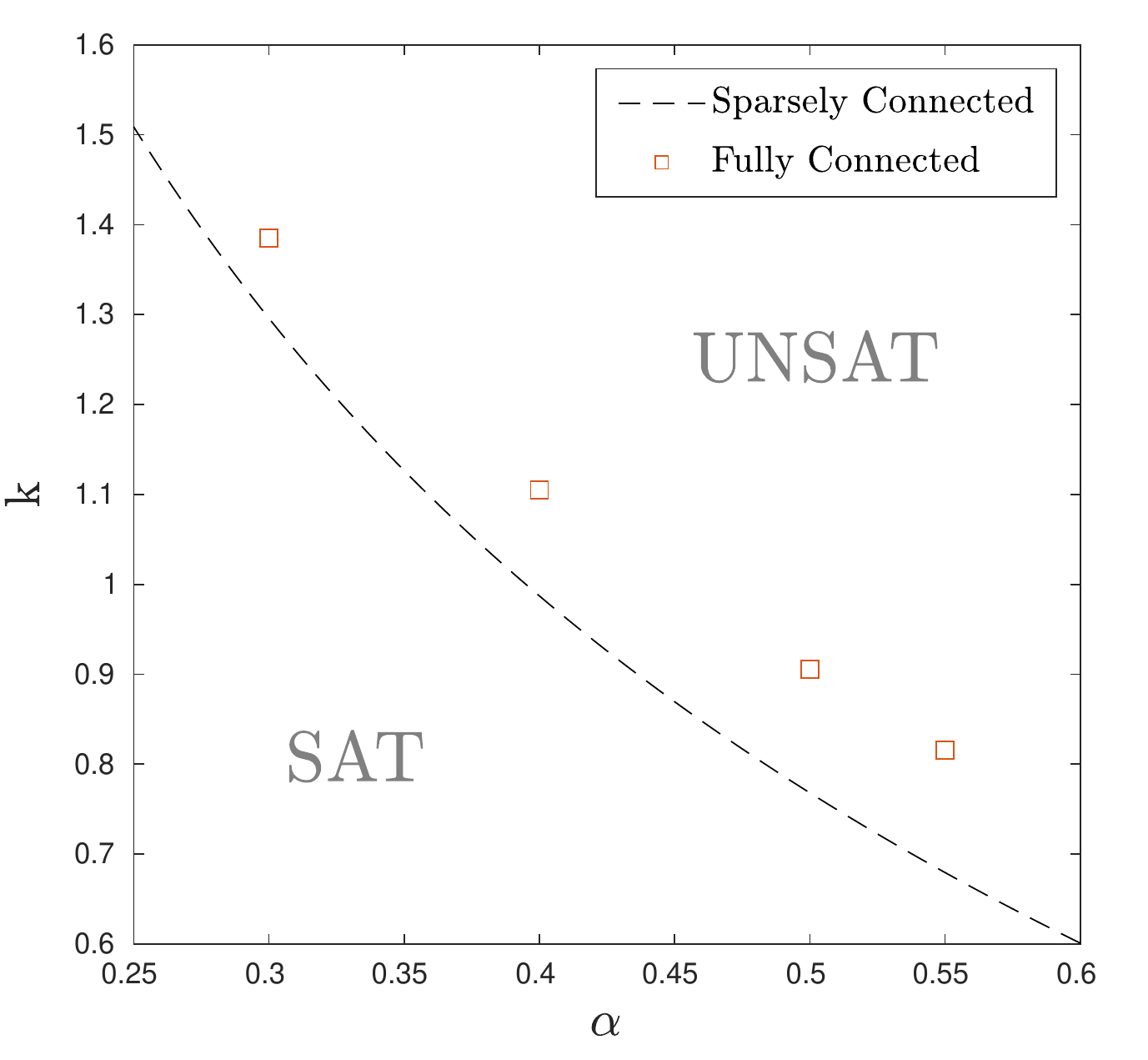}
    \caption{Phase diagram of the linear symmetric perceptron in the plane defined by the pattern density $\alpha$ and the stability parameter $k$. The dashed line is the analytical result for $k_{max}(\alpha)$ obtained for slightly diluted networks~\cite{Gardner89}. Squares show numerical results for $k_{max}(\alpha)$ in a fully connected model at $\alpha \in \{0.3, 0.4, 0.5, 0.55\}$. Simulations have been run at different sizes of the network to measure the probability for the algorithm to converge before $10^3$ steps of the training (hence providing a lower bound to the actual value of stability). A standard finite size scaling analysis~\cite{barber1983domb,altarelli2009review} has been used to extrapolate the value of $k_{max}(\alpha)$ to the thermodynamic limit.}
\label{fig:sat_unsat}           
\end{figure}

In the symmetric case,
the function $k_{max}(\alpha)$ has been determined analytically~\cite{Gardner89} for slightly diluted recurrent networks, i.e. networks with an average connectivity scaling as $\log N$. Numerical results from the study of the algorithm defined in eq.~(\ref{eq:gardner_update_rule}) on networks that are both fully connected and fully symmetric suggest that, for the same degree of symmetry, $k_{max}$ at a given $\alpha$ is located slightly above the one predicted by \cite{Gardner89}. This finding, discussed in Fig.~\ref{fig:sat_unsat}, suggests to reconsider previous interesting analyses~\cite{theumann} and opens the road to further investigations of the critical capacity as a function of the network connectivity~\cite{maurofabian}. 

\begin{figure}
\includegraphics[width=.5\textwidth]{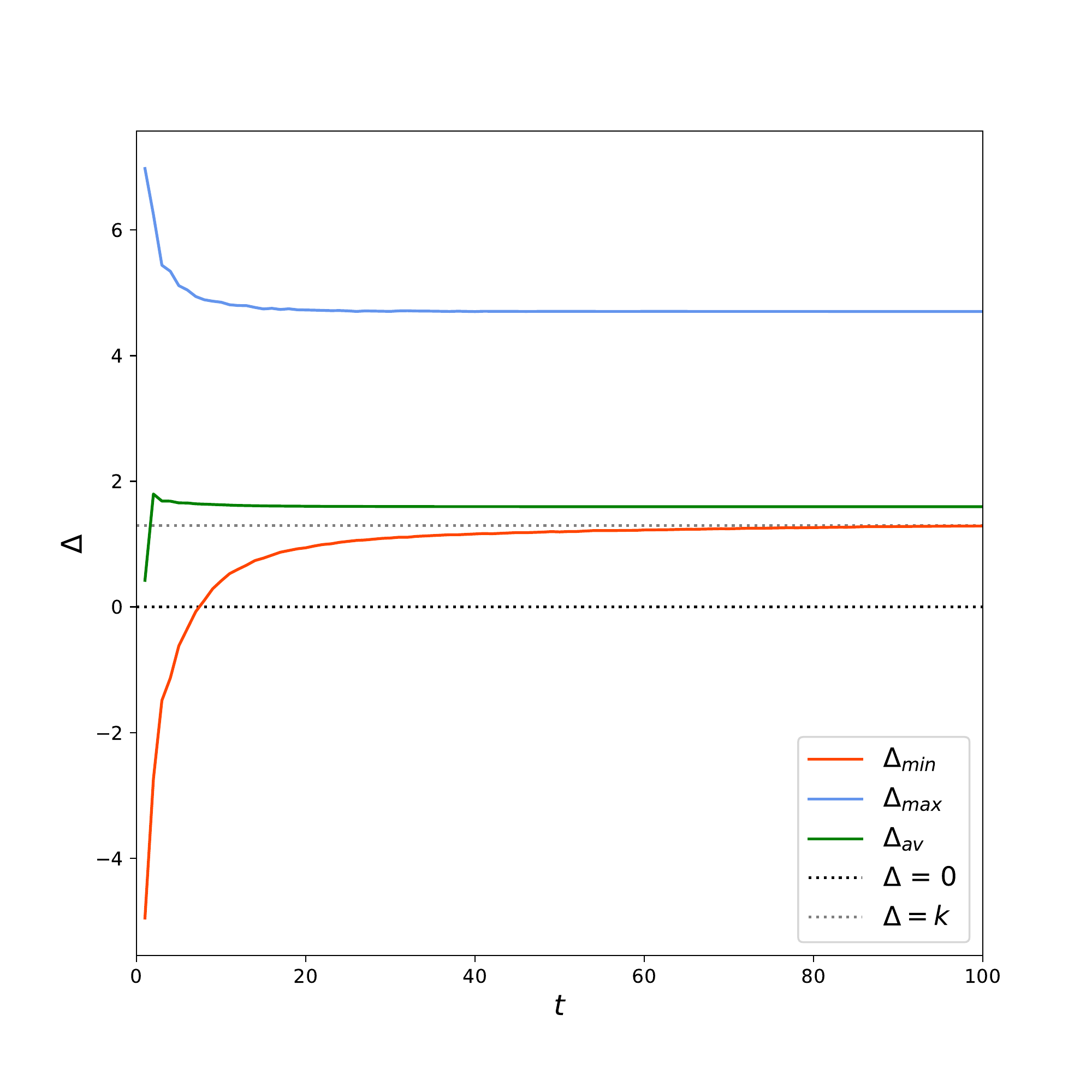}
    \caption{Values of the minimal stability $\Delta_{min}$ (orange), maximal stability $\Delta_{max}$ (blue) and average stability $\Delta_{av}$ (green) are plotted as a function of the number of iterations $t$ of the symmetric perceptron algorithm, averaged over $50$ realizations of the memories, for $N = 800$, $\alpha = 0.3$, $\lambda = 1$. The black dotted line represents the zero-stability threshold that must be overcome by $\Delta_{min}$ to have all memories perfectly recalled. The gray dotted line represents the minimum stability required by the algorithm to be reached at convergence, i.e. $\Delta = k$.}
\label{fig:stability_SP}           
\end{figure}

Because we are analyzing the close relationship between the symmetric perceptron and Hebbian unlearning (see Sec. \ref{sec:2b}), we always initialize the couplings according to Hebb's rule, Eq.~(\ref{eq:hebbian_learning}), keeping in mind that results for $k=k_{max}(\alpha)$ should not depend on this choice. 
In Fig.~\ref{fig:stability_SP} we plot the evolution with the number of steps $t$ of $\Delta_{min}:=\min_{i,\mu} \{\Delta_i^\mu\}$, $\Delta_{max}:=\max_{i,\mu} \{\Delta_i^\mu\}$ and $\Delta_{av}:=1/(PN)\sum_{i,\mu}\Delta_i^\mu$, each averaged over many random realizations of the memories~$\vec{\xi}$. One sees that the SP algorithm, while slightly reducing $\Delta_{av}$ and $\Delta_{max}$, increases the value of $\Delta_{min}$ from negative values (i.e. not all memories are stable) to positive values (i.e. all memories are stable) and up to the prescribed threshold $k$. The same profile of the stabilities is obtained at different choices of the control parameters, when $k \leq k_{max}(\alpha)$.  

\subsection{\label{sec:2b}Hebbian unlearning}

A very interesting proposal to increase the performance of the Hopfield model consists of modifying the Hebbian learning rule defined in Eq.~(\ref{eq:hebbian_learning}) by adding to it a ``dreaming" procedure \cite{Hop83,vanhemmenHebbianLearningIts1998,Agliari}. This goes as follows: initialize the coupling matrix according to Eq.~(\ref{eq:hebbian_learning}), and then repeat $D$ times the following steps:
\begin{itemize}
    \item Initialize the neurons to a random state, and follow the dynamics until it converges to a fixed point $\vec{\sigma}^*$.
    \item Modify the coupling matrix according to 
    \begin{equation}
     J_{ij}\to J_{ij}-\frac{\epsilon}{N}\sigma^*_i\sigma^*_j, \quad J_{ii}=0 \ .
     \label{eq:unlearning_algorithm}
     \end{equation}
\end{itemize}
This algorithm is unsupervised, in the sense that it does not need to be provided explicitly with the memories $\{\vec{\xi}^\mu\}$, and only exploits the information encoded in Hebb's learning rule \cref{eq:hebbian_learning}. It is easy to see that at each step the energy of the configuration $\vec{\sigma}^*$ reached by the algorithm is increased, making it less stable. Since every memory is surrounded by many spurious local energy minima, the overall effect is a smoothing of the energy landscape around the attractors correlated to the memories, by destabilizing other attractors. This procedure is often referred to as Hebbian unlearning and each iteration of the algorithm is referred to as a \textit{dream}. The total number of dreams $D$ is a parameter of the algorithm, and must be chosen as to maximize the recognition performance at a given load $\alpha$. It has been shown that this procedure improves the performance of the model in two ways \cite{VH90,VH92,VH94,Klein}: on the one hand, the critical memory load is increased up to $\alpha_c^{HU}\sim 0.6$. On the other hand, the overlap between the attractors and the memories goes to one, meaning that memories become real fixed points of the dynamics. 

\begin{figure}
\centering
\includegraphics[width=.5\textwidth]{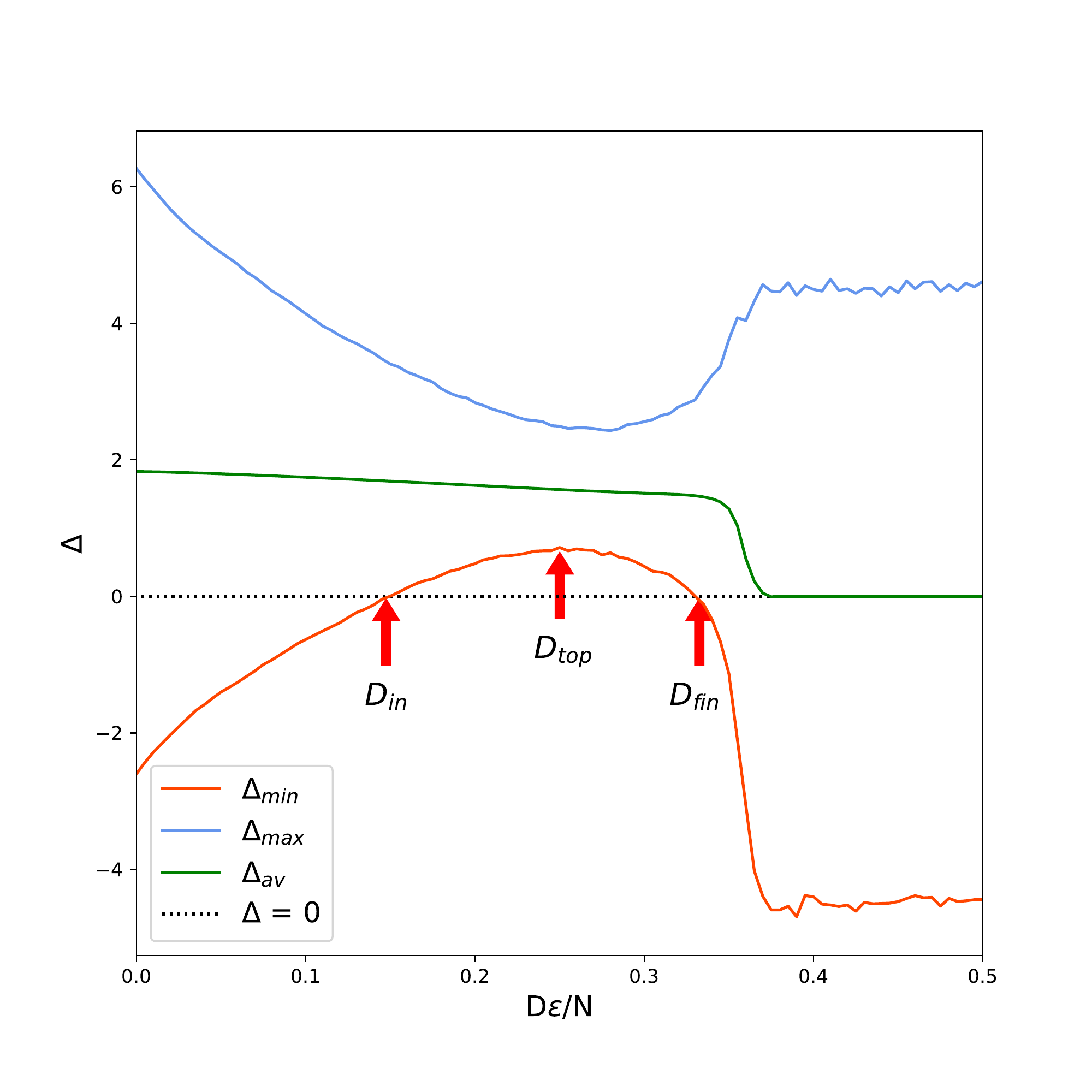}
    \caption{Values of the minimal stability $\Delta_{min}$ (orange), maximal stability $\Delta_{max}$ (blue) and average stability $\Delta_{av}$ (green) computed during Hebbian unlearning, averaged over $50$ realizations of the memories at $N = 800$, $\alpha = 0.3$, $\epsilon = 10^{-2}$. The black dotted line represents the zero-stability threshold to be overcome by $\Delta_{min}$ to have all memories perfectly recalled. We denote the corresponding value of the number of dreams $D$ by $D_{in}$. $D_{top}$ is for the point where the algorithm reaches the maximum value of  $\Delta_{min}$, while $D_{fin}$ is the end of the perfect classification regime of the network. We used red arrows to point at $D_{in}$, $D_{top}$ and $D_{fin}$.}
\label{fig:stability_HU}           
\end{figure}

This last fact, which is especially remarkable because the dreaming procedure is unsupervised, led us to try and characterize the performance of the algorithm in terms of the stabilities $\Delta_i^\mu$ introduced in Sec.~\ref{sec:2a}. This approach has already been attempted~\cite{Horas}, and our analysis pushes it further and reveals new unexpected features.  In fig.~\ref{fig:stability_HU} we show the typical behavior of $\Delta_{min},\,\Delta_{av}$ and $\Delta_{max}$ as the unlearning procedure unfolds. The horizontal axis represents the number of steps $D$ performed by the algorithm, rescaled by $\epsilon/N$. The reason for this choice will become clear later. Focusing on $\Delta_{min}$, we can see a non-monotonic behavior: the minimal stability grows to positive values, peaks at some value $D=D_{top}$ and then decreases back to negative values. Between $D_{in}$ and $D_{fin}$ every stability is positive or, equivalently, every memory is a fixed point for the dynamics. As we increase $\alpha$, the interval $[D_{in}, D_{fin}]$ shrinks, and the height of the peak at $D_{top}$ lowers, until we reach a critical load $\alpha_c$, above which $\Delta_{min}$ never goes above zero. Collecting data for networks of size $N=300,\,400,\,500,\,600,\,800$ and different values of $\alpha$ and $\epsilon$ it is possible to extrapolate the position of $D_{in},\,D_{top}$ and $D_{fin}$ as a function of $\alpha,\,\epsilon$ and $N$, as well as the critical capacity $\alpha_c$. By fitting the data with respect to the model parameters we found that the number of dreams is in every case linear in $N$ and $1/\epsilon$. Moreover, $D_{top}$ also depends linearly on $\alpha$. At the critical capacity,
$$\alpha_c^{HU}=0.589\pm0.003 \ ,$$
the value of $\Delta_{min}(D_{top})$ approaches zero. 
The value of the critical capacity $\alpha_c^{HU}$ as well as the linear dependence of $D_{in}$, $D_{top}$, $D_{fin}$ on $N/\epsilon$ are consistent with the past literature~\cite{VH92, VH90, Horas}. 

Because the $\Delta_{min}(D)$ curve is quadratic around $D_{top}$, as illustrated in
Fig.~\ref{fig:stability_HU},
$D_{in}$ and $D_{fin}$ both tend to $D_{top}$ at the critical capacity  with a critical exponent $1/2$. 
The resulting scaling relations are: 
\begin{equation}
    \label{eq:top}
    D_{top}(\epsilon,\alpha,N)=\frac{N}{\epsilon}(a\cdot\alpha+b) \ ,
\end{equation}
\begin{equation}
    \label{eq:in}
    D_{in}(\epsilon,\alpha,N)=D_{top}-\frac{N}{\epsilon}(c\cdot\alpha+d)^{1/2} \ ,
\end{equation}
\begin{equation}
    \label{eq:fin}
    D_{fin}(\epsilon,\alpha,N)=D_{top}+\frac{N}{\epsilon}(e\cdot\alpha+f)^{1/2} \ ,
\end{equation}
with
$$ a=1.02\pm0.02 \ , \hspace{0.5cm} b=-0.05\pm0.01 \ ,$$
$$ c=-0.039\pm0.003 \ ,\hspace{0.5cm} d=0.023\pm0.002 \ ,$$
$$ e=-0.022\pm0.001 \ ,\hspace{0.5cm} f=0.013\pm0.001 \ .$$
All the statistical errors have been evaluated using the jackknife method.

\section{\label{sec:3}Basins of Attraction}
\begin{figure}
\centering
\includegraphics[width=.5\textwidth]{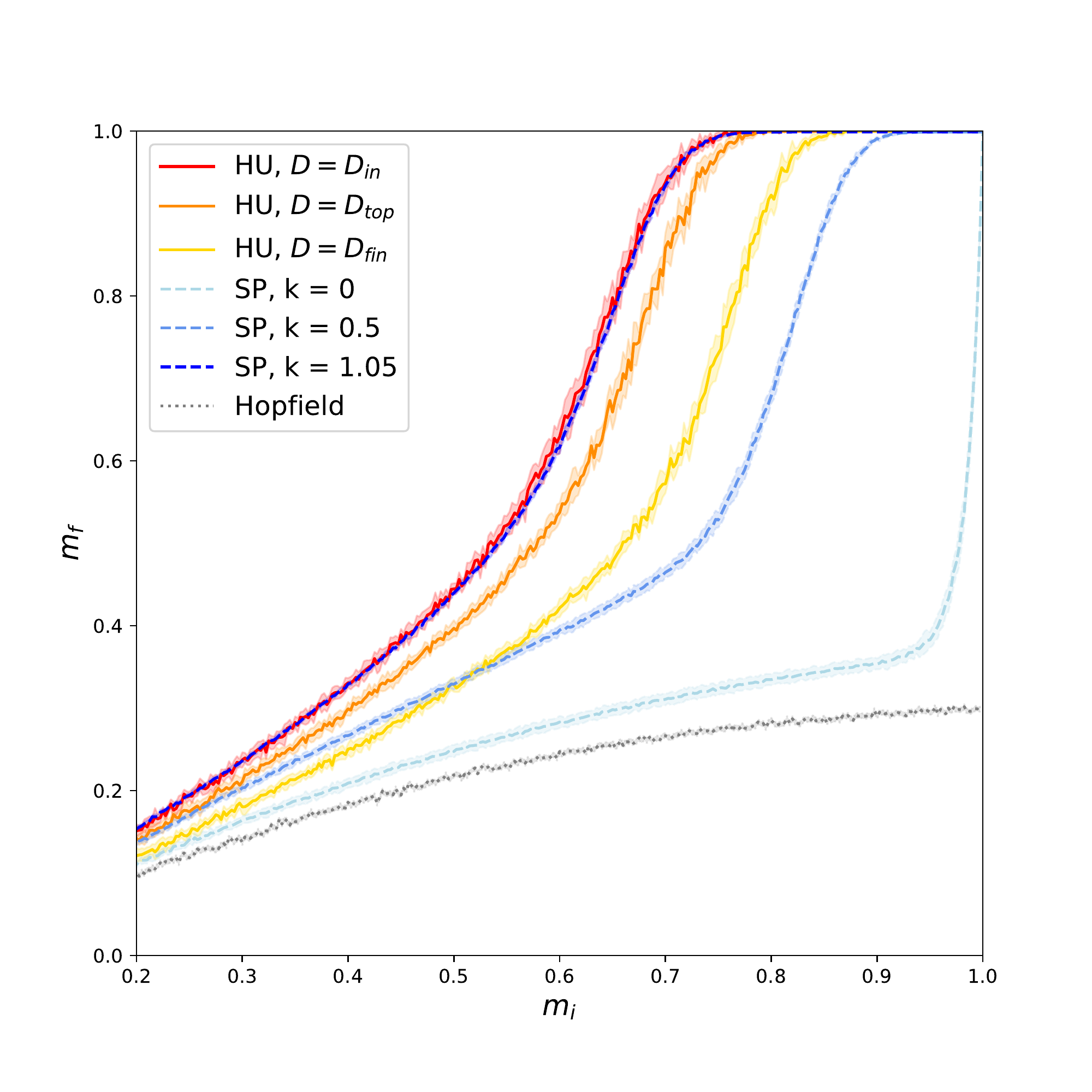}
    \caption{The average size of basins of attraction at $N = 800$, $\alpha = 0.4$ for both symmetric perceptron (SP) and Hebbian unlearning (HU), averaged over several realizations of the disorder. The colored area around each curve represents the statistical errors. Continuous lines are for basins at $D_{in}, D_{top}$ and $D_{fin}$ for HU with $\epsilon = 10^{-2}$. Dashed lines are for three values of $k$, including the $k \simeq k_{max}$, used for SP with $\lambda = 1$. The gray dotted line represents the performance of the Hopfield model at the same value of $\alpha$. Notice that attraction basins of the SP and HU almost coincide when the two algorithms operate in their optimal regime, namely $D=D_{in}$ and $k\simeq k_{max}$.}
\label{fig:mf_mi}           
\end{figure}

We have also compared the performance of SP and HU by measuring the shape of the basins of attraction around each memory. This is done by initializing the network at some fixed distance $m_i$ from one of the memories $\vec{\xi}^\mu$ and following the dynamics until convergence to a fixed point $\vec{\sigma}^*$ is reached. Then, we measure the overlap between $\vec{\sigma}^*$ and $\vec{\xi}^\mu$, averaged over many realizations of the memories:
\begin{equation}
    m_f=\frac{1}{N}\sum_{i=1}^N\left\langle\xi^\mu_i\sigma^*_i\right\rangle.
\end{equation}
In Fig.~\ref{fig:mf_mi} we plot $m_f$ as a function of $m_i$. Colored dashed curves refer to SP for different values of $k$, up to the highest $k$ that allows the algorithm to converge in $O(10^3)$ iterations. This slight underestimations of the real $k_{max}(\alpha)$ bares very little consequences to our results. Related to this, we underline the importance of the choice of $\lambda$ at a given value of $N$. This is another crucial topic that is rarely discussed in the literature. Higher values of $\lambda$ imply larger learning steps, while smaller values are associated to a finer exploration of space of coupling matrices during training. It is observed that the algorithm, operating at $\lambda = 1,10^{-1}, 10^{-2}$, converges to almost identical matrices already when $k$ is equal to the maximal stability for diluted networks~\cite{Gardner89} that, according to Fig.~\ref{fig:sat_unsat}, is slightly lower than the actual $k_{max}$.
This suggests that the final state lies very closely to the unique optimal solution even when we are not exactly at $k_{max}$. Hence, no significant changes are expected in our numerical results when $k$ is pushed further towards its maximal value. On the other hand, when $\lambda$ assumes smaller values, i.e. $\lambda = 10^{-3}, 10^{-4}, 10^{-5}$, basins are observed to be smaller in size and the volume of solutions is larger, indicating that the final state remains farer from the maximal performance. In order to recover the numerical results obtained at a larger $\lambda$, one needs to progressively increase $k$ to values that are difficult to reach numerically. As a result, the choice of $\lambda = 1$ in this section seems to us well justified to reproduce the optimal performance of the symmetric perceptron at $k \simeq k_{max}$. 

Consistently with the literature, we find that increasing the stability leads to an increase of $m_f$ at fixed $m_i$~\cite{Forrest}. In particular, when the stability is equal to zero, the memories, albeit being fixed point of the dynamics, have zero basin of attraction, as indicated by the very low values of $m_f$ for $m_i\neq 1$. The gray dashed line at the bottom of Fig.~\ref{fig:mf_mi} refers to the Hopfield model without dreaming: since $\alpha>0.14$ the model does not learn. The colored continuous lines refer to Hebbian unlearning, for different amounts of dreaming. More specifically, we measured the performance of the model for the three values $D=D_{in},\,D_{top},$ and $D_{fin}$ defined in Sec.~\ref{sec:2b}. It is clear how dreaming improves the performance of the network, and we found that the performance is not maximized at $D=D_{top}$ as one could expect~\cite{Horas}, but at $D=D_{in}$, where the requirement for perfect retrieval of the memories is satisfied with zero margin. 

\begin{figure}
\centering
\includegraphics[width=.5\textwidth]{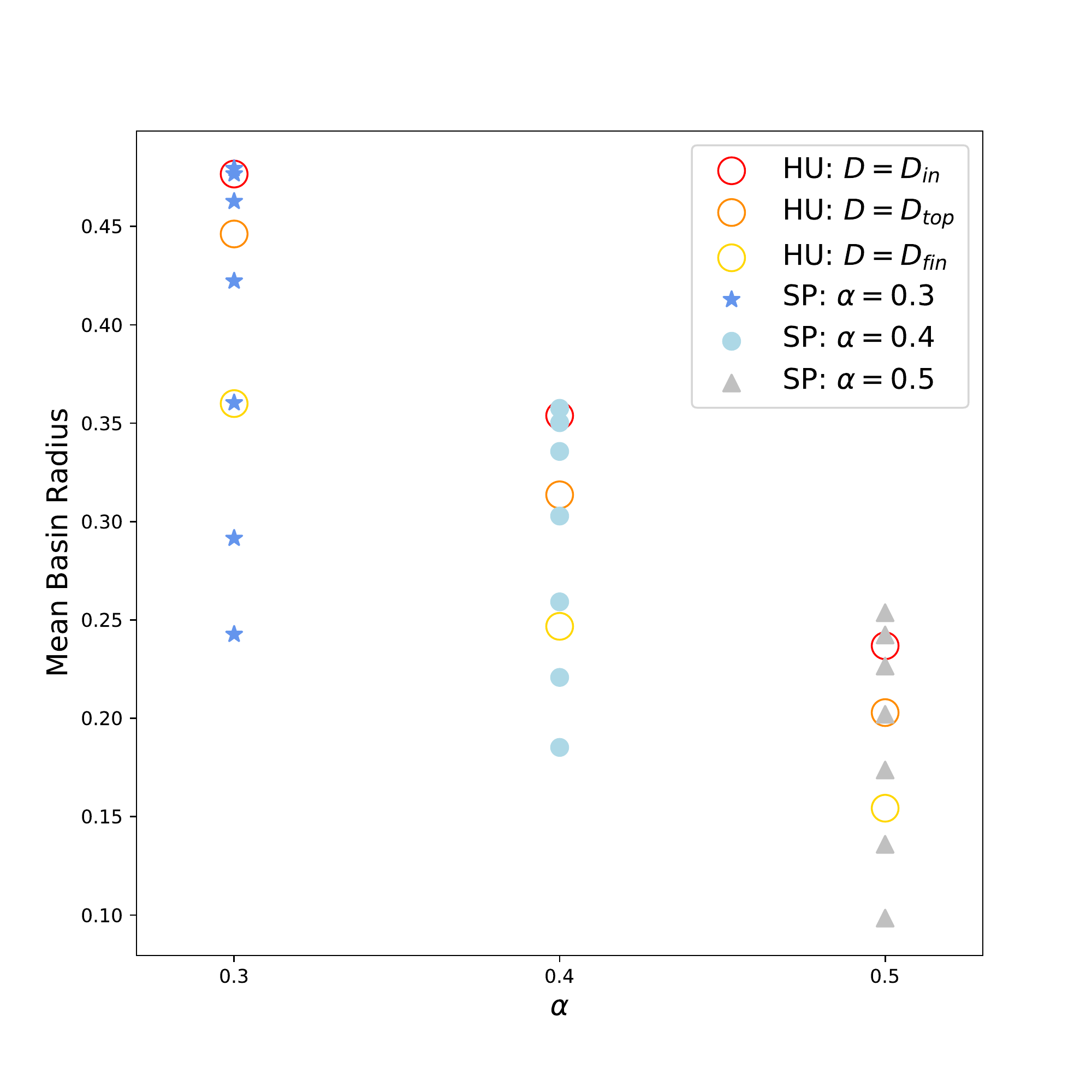}
    \caption{ Mean attraction basin radius for symmetric perceptron (SP) and Hebbian unlearning (HU) measured as in~\cite{Forrest} and extrapolated to $N\to\infty$, for $\alpha = 0.3, 0.4, 0.5$ and $\lambda = 1$. Points for the SP correspond to the following values of $k$: $\alpha = 0.3\rightarrow k\in\{0.4, 0.5, 0.7, 0.9, 1.1, 1.296, 1.32\}$; $\alpha = 0.4\rightarrow k\in\{0.4, 0.5, 0.6, 0.75, 0.9, 0.988, 1.05\}$; $\alpha = 0.5\rightarrow k\in\{0.3, 0.4, 0.5, 0.6, 0.7, 0.768, 0.85\}$. Error bars are smaller than the symbols size. } 
\label{fig:basins_Ninf}           
\end{figure}

We also found that the performance of Hebbian unlearning at $D=D_{in}$ and the one of the SP at $k\simeq k_{max}$ are indistinguishable within our numerical resolution. This is a remarkable fact, since the two algorithms have a radically different structure: the SP algorithm is supervised, i.e. it needs to have access at every step to all the memories that the network needs to memorize, while the HU is not, and only exploits the topology of the spurious states generated by Hebb's prescription in Eq.~\eqref{eq:hebbian_learning}. These findings are robust to change in the load $\alpha$ and to finite size effects, as illustrated in Fig.~\ref{fig:basins_Ninf}. The mean basin radius at finite $N$ is defined as $1 - m_i$, selecting the value of $m_i$ below which more than $30\%$ of the memories are reconstructed with more then $5\%$ error. The dots represent our extrapolation of this quantity to the limit $N\to\infty$, for different values of $\alpha$. The lower dots relative to the SP correspond to $k<k_{max}$, and the value of the mean basin radius gets higher as $k$ is increased up to $k \simeq k_{max}$. Again, one can see that even in the thermodynamic limit, our simulations suggest that in their optimal regime the two algorithms perform essentially in the same way.

\section{\label{sec:4}Space of Interactions}

One way to visualize the solutions of the optimization problem, and the way these solutions are reached by means of the algorithm, is to exploit the space of interactions as conceived by Gardner~\cite{Gardner88}. Consider a spherical surface in $N(N-1)/2 - 1$ dimensions where each point is a vector composed by the off-diagonal elements $J_{j > i}$ of the connectivity matrix normalized by their standard deviation. These position vectors hence will be
\begin{equation}
    \label{eq:r}
    \vec{r} = \vec{J}/\sigma_J\ ,
\end{equation}
with
\begin{equation}
    \label{eq:sigma}
    \sigma_J = \sqrt{\frac{2}{N(N-1)}\sum_{i < j}^{1,N} J_{ij}^2}\ .
\end{equation}
For what concerns the SP, after fixing the value of $\alpha$ and a set of patterns, one can imagine the sphere as composed by an UNSAT and a SAT region. These regions are connected sub-spaces of the original sphere, so that one can go from a matrix to another one in a continuous fashion. The SAT region contains the point relative to the unique solution at $k = k_{max}(\alpha)$.

We now define an \textit{overlap} parameter quantifying the covariance of two generic symmetric matrices $J_{ij}$ and $U_{ij}$
\begin{equation}
    \label{eq:q}
    q = \frac{2}{N(N-1)} \sum_{i < j}^{1,N}\langle \frac{J_{ij}U_{ij}}{\sigma_J \sigma_U} \rangle \ ,
\end{equation}
where $\langle \cdot \rangle$ is the average over the disorder.

\subsection{\label{sec:4a}Final States}

\begin{figure*}
\centering
\includegraphics[width=.5\textwidth]{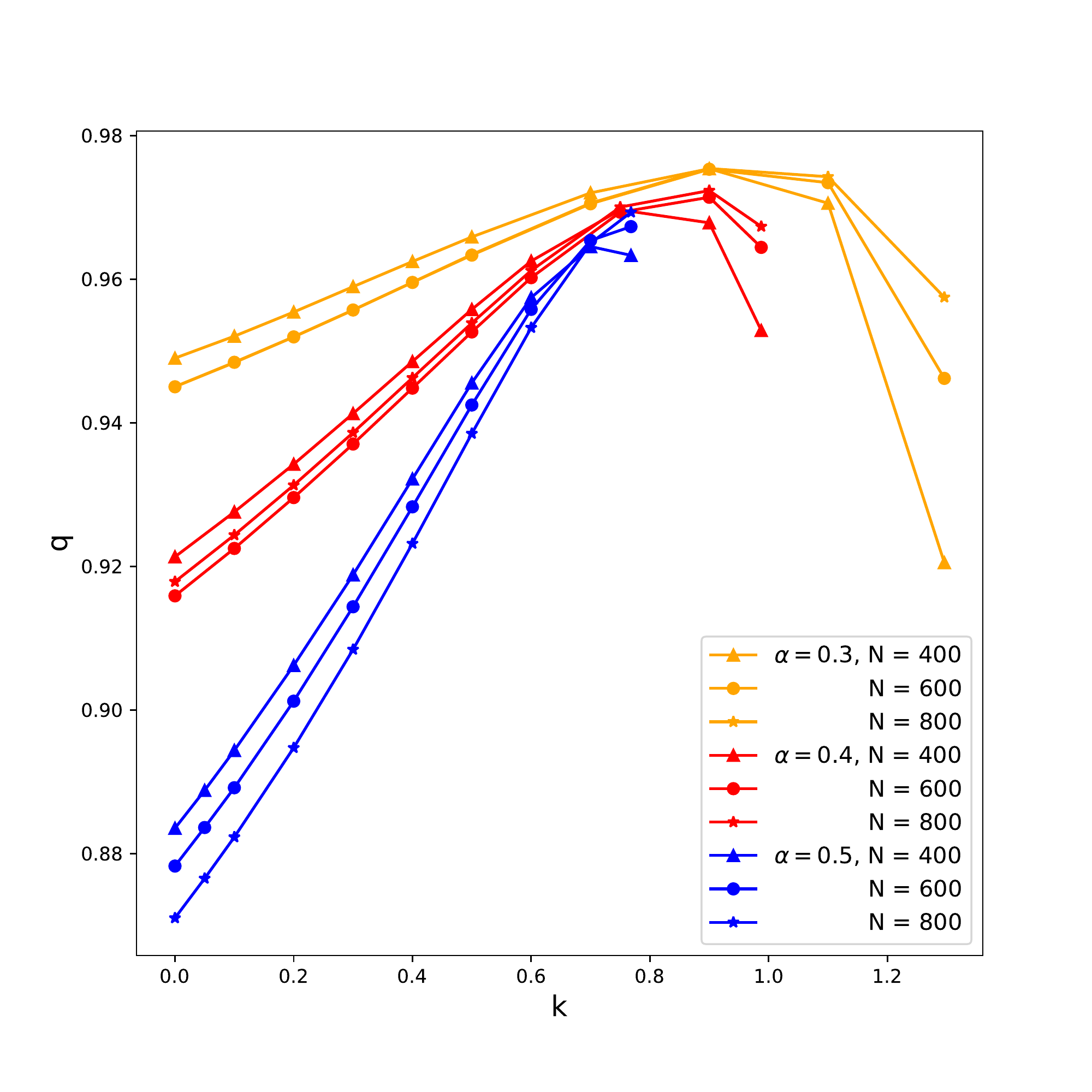}\hfill
\includegraphics[width=.5\textwidth]{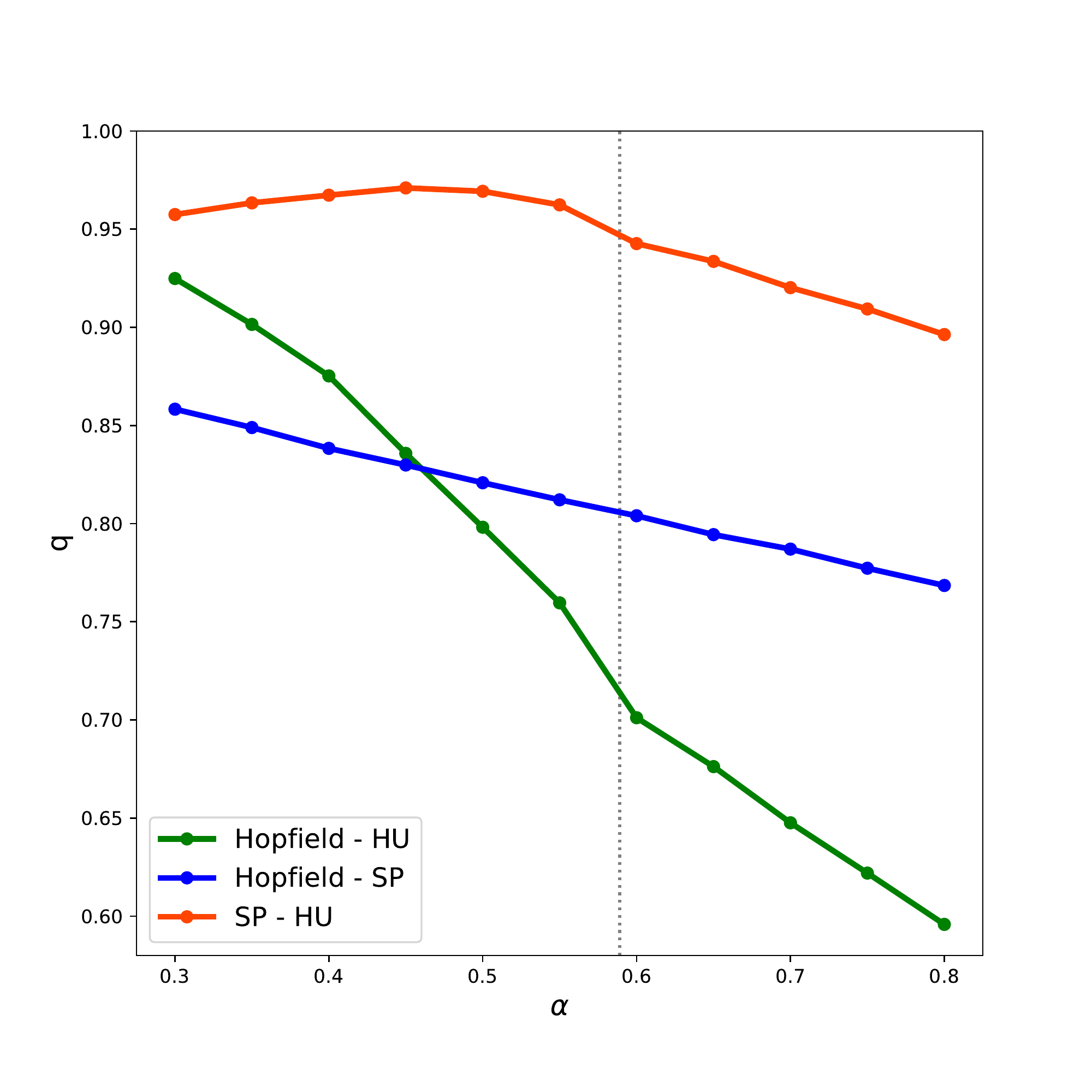}\hfill
    \caption{Left: Overlap $q$ between the final states of Hebbian unlearning (HU) with $\epsilon = 10^{-2}$ at $D = D_{in}$ and symmetric perceptron (SP) with $\lambda = 1$ having reached a stability $k$. Measures are for different values of $N$ and $\alpha$ and points represent the mean computed over $5$ realizations of the disorder. Error bars are smaller than the data symbol. Values of $k$ range from $0$ to slightly below $k_{max}(\alpha)$. For each $\alpha$, $q$ peaks around $k_{max}(\alpha)$, indicating that the two algorithms converge to coupling matrices which are closest near to the value $k=k_{max}(\alpha)$. Right: Overlap $q$ as a function of $\alpha$ at $N = 800$. Points represent the mean of $10$ realizations of the disorder and error bars are  smaller than the data symbol. The orange symbols correspond to the overlap between the final states of SP with $\lambda = 1,\,k\simeq k_{max}(\alpha)$ and HU with $\epsilon = 10^{-2}$. For HU we chose $D = D_{in}$ for $\alpha<\alpha_c$, while for $\alpha>\alpha_c$ we chose $D=D_{top}$ ($\alpha_c$ is represented by the gray dotted line). The overlap between the initial Hebbian matrix and the final state of the HU (green) or SP (blue) with the same choice of the parameters is also shown. While in both algorithms the distance between initial and final matrix increases as $\alpha$ is increased, the distance between the final points remains small up to $\alpha_c$.}
\label{fig:overlap}           
\end{figure*}

We first evaluate the final points where the two algorithms converge in the space of interactions. Hebbian Unlearning is stopped at $D = D_{in}$ as that is the relevant amount of $\textit{dreams}$ identified in Sec. \ref{sec:2b}. The Symmetric Perceptron is run at $\lambda = 1$. Fig.~\ref{fig:overlap}(left) displays the overlap between the resulting matrices when the SP is performed at different values of $k$ before reaching $k_{max}(\alpha)$. The plot shows that $q$ increases with $k$, suggesting that HU pushes the system to the same region of solutions where the SP converges when $k$ is close to $k_{max}$. Finite size effects evidently appear near the abrupt transition from SAT to UNSAT, but the increase of $q$ with the size of the network suggests that the maximum overlap might be associated to the maximal stabilities when $N$ becomes large enough. 

The plot of $q$ as a function of $\alpha$, see Fig.~\ref{fig:overlap}(right), shows how the distance between the final points and the initial Hebbian matrix increases when the number of memories becomes larger, while the distance between the two final points remains small and stable for $\alpha < \alpha_c^{HU}$. 

\subsection{\label{sec:4b}Learning Paths and Gradients}
By comparing the final states of convergence as done in Sec.~\ref{sec:4a} we conclude that two networks, starting from the same initial matrix, end up in very similar configurations of the couplings $J_{ij}$. Now we analyze  the whole trajectory traced by the two algorithms in the space of interactions. 

\begin{figure*}
\centering
\includegraphics[width=.55\textwidth]{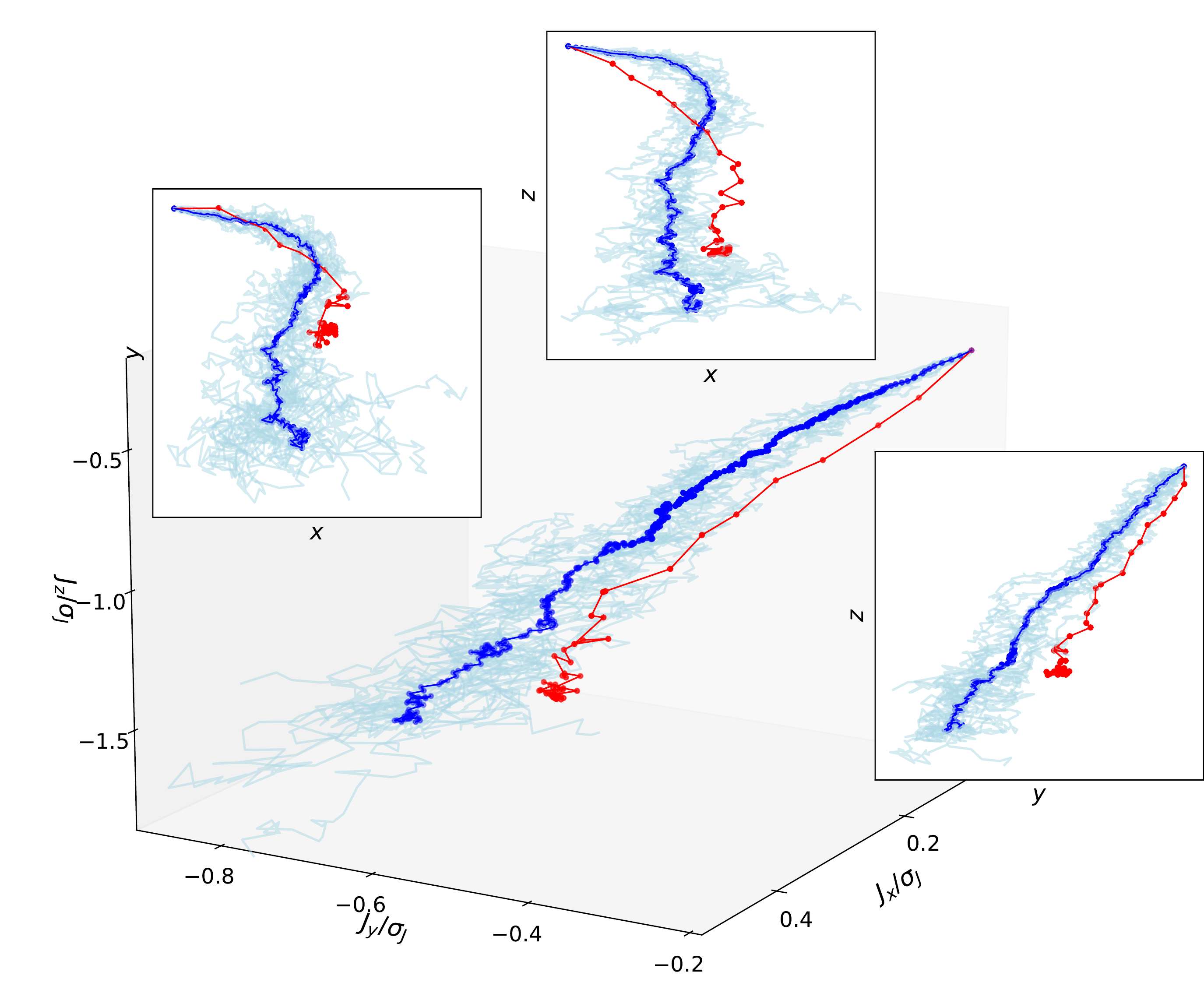}\hfill
\includegraphics[width=.45\textwidth]{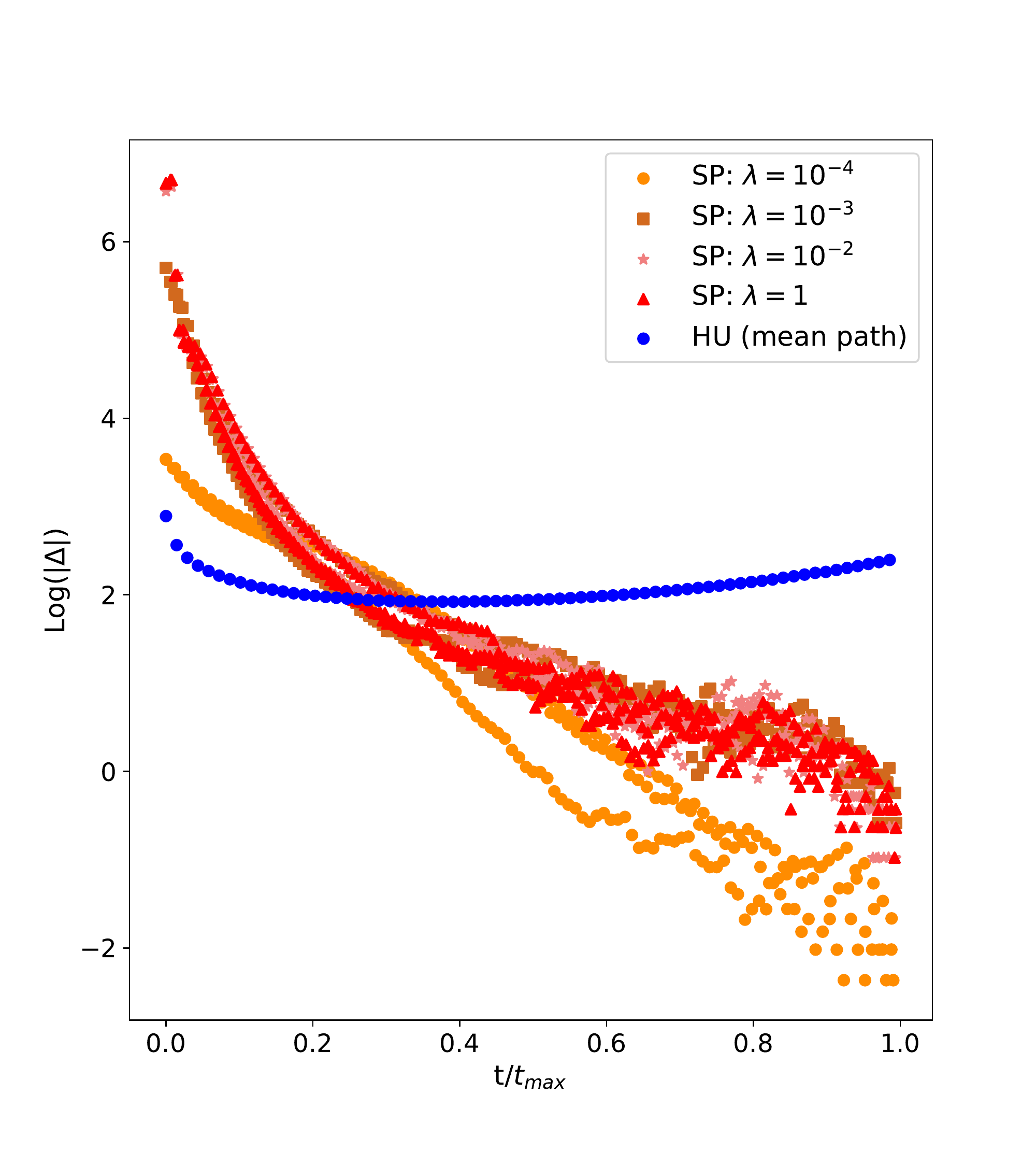}\hfill
    \caption{Left: 3-dimensional projection of the trajectories followed by the system  in the space of interactions during the dynamics of Hebbian unlearning (HU) and symmetric perceptron (SP). Numerical measurements have been taken for one sample at $N = 800$ and $\alpha = 0.55$, $\epsilon = 10^{-2}$, $\lambda = 10^{-4}$. $10$ trajectories of the HU are drawn in light blue, while the average unlearning path is in blue. The path followed by the SP is depicted in red. Points represent different steps of the algorithms. HU has been resampled at regular intervals along the trajectory for simplicity of the data analysis. Right: Absolute value of the variation $\Delta \vec{J}$ in logarithmic scale as a function of the normalized time scale $t/t_{max}$ where $t_{max}$ is the maximum number of steps reached by the algorithm in a given sample. Numerical measurements are for one sample at $N = 800$ and $\alpha = 0.55$, $\epsilon = 10^{-2}$, $\lambda = 1, 10^{-2}, 10^{-4}$.  Three samples were simulated for the SP and one sample for the HU.}
\label{fig:trajectories}           
\end{figure*}

We set $\alpha = 0.55$, so that the overlap between the initial and the final state is small enough, i.e. they are distant on the sphere, $N = 800$, $\lambda = 10^{-4}$ and $k$ close to $k_{max}$ in one single sample. The choice of a small value of the learning rate $\lambda$ allows to trace a continuous path in the space of the interactions. HU is run choosing $D = D_{in}$ for $10$ samples in total. Fig.~\ref{fig:trajectories}(left) reports the projection of the resulting trajectories in the space of $J$ along three randomly chosen directions. The plot shows that the two algorithms explore the same region of the space of interactions, proceeding along a similar direction. We also observe that the convergence velocities of the two algorithms are very different. Indicating with $t$ the time steps for both processes, Fig.~\ref{fig:trajectories}(right) shows the logarithm of the absolute value of the \textit{variation} of vector $\vec{J}$, defined as 
\begin{equation}
    \label{eq:gradient}
    {\Delta}\vec{J}^{(t)} = \vec{J}^{(t+1)}/\sigma_{J}^{(t+1)} - \vec{J}^{(t)}/\sigma_{J}^{(t)} \ .
\end{equation}
The direction of this vector coincides with the one of the gradient followed by the algorithm in the space of interactions at a given time step.  

While the convergence speed of the HU does not significantly vary, the SP shows, at any scale of $\lambda$, an acceleration in time that resembles an exponential law. In other words, while HU explores the space of interactions nearly uniformly in speed, the SP takes about $15 \div 20$ time steps to reach a smaller condensed region where it gets confined until convergence.

\begin{figure*}
\centering
\includegraphics[width=.33\textwidth]{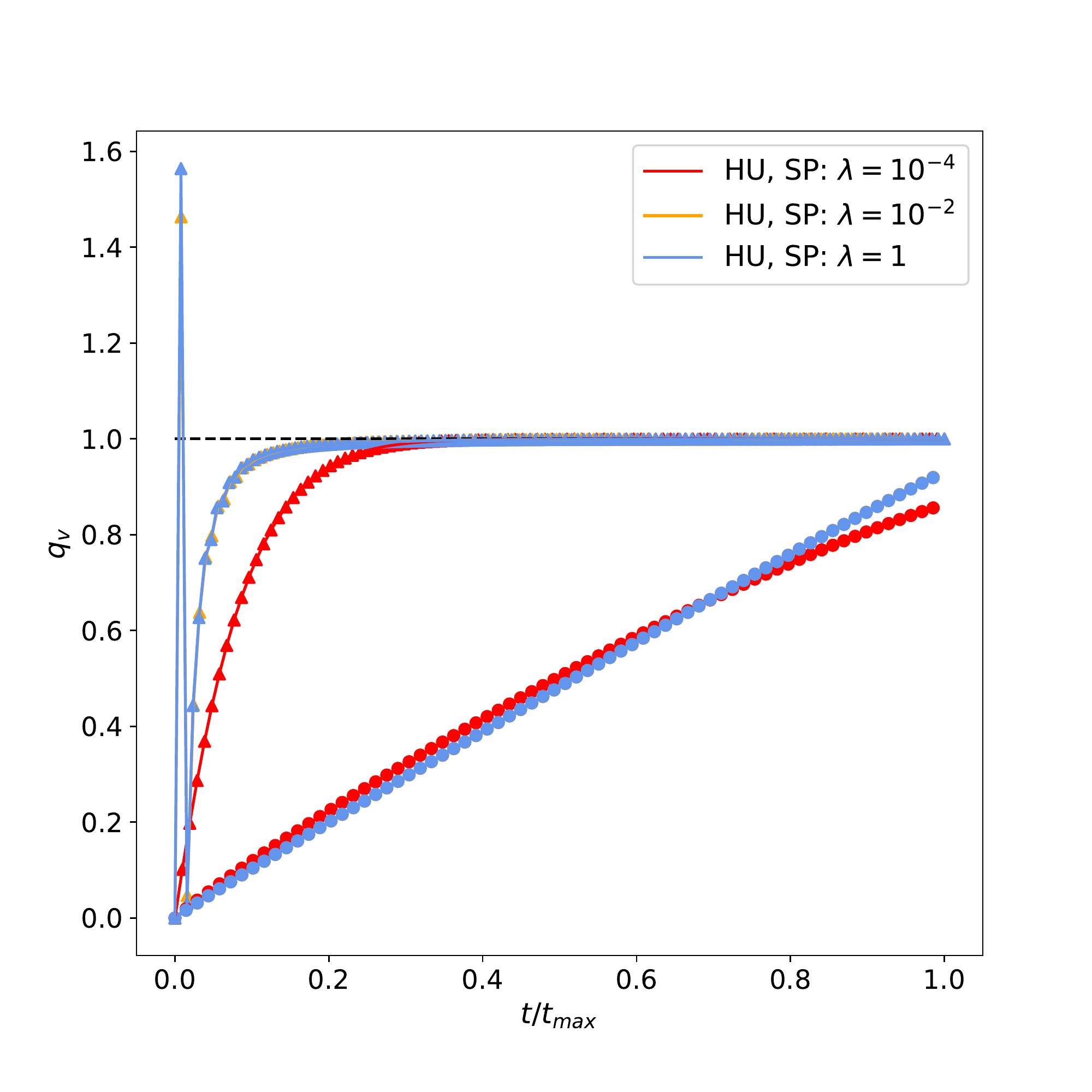}\hfill
\includegraphics[width=.33\textwidth]{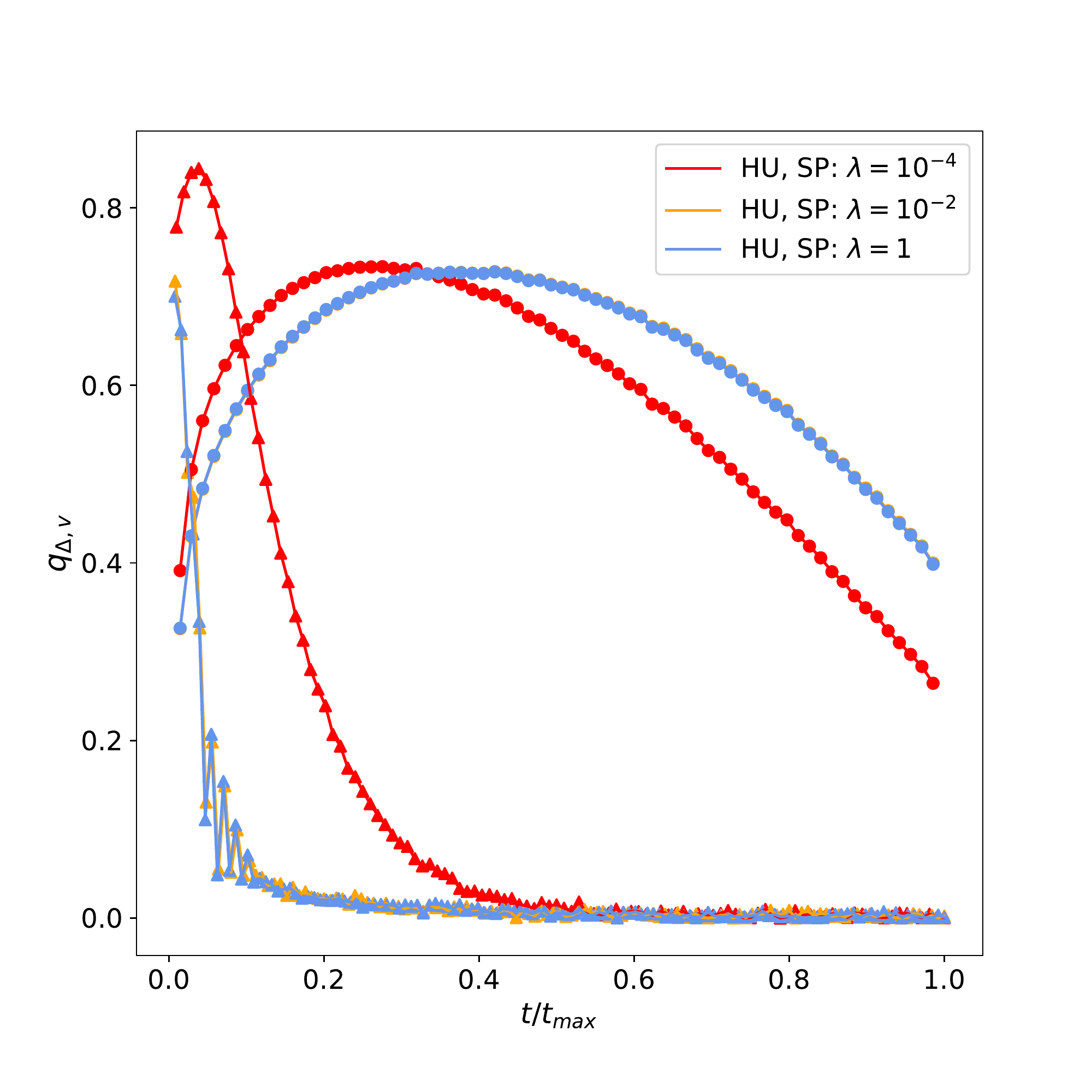}\hfill
\includegraphics[width=.33\textwidth]{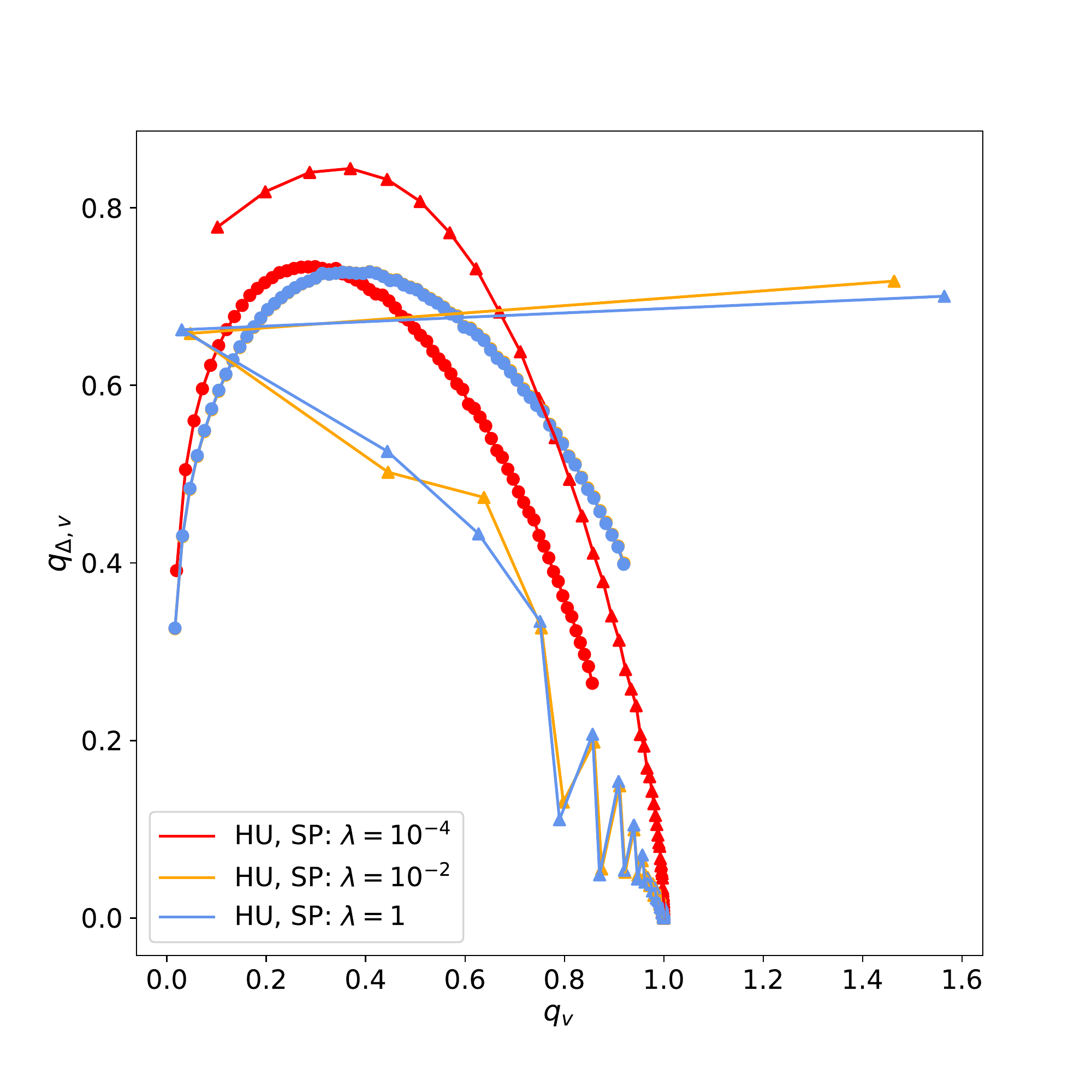}\hfill
    \caption{Left: $q_v$, angular distance of the trajectory from the reference direction $\hat{v}$,  as a function of time. Center: $q_{\Delta,v}$, projection of the variation along the direction $\hat{v}$, as a function of time. Right: $q_{\Delta,v}$ as a function of $q_v$. Numerical measurements are collected from one single sample at $N = 800$ and $\alpha = 0.55$ at $\epsilon = 10^{-2}$ for the Hebbian unlearning (HU) (\textit{circles}) and different values of $\lambda$ for the symmetric perceptron (SP) (\textit{triangles}).}
\label{fig:gradients}           
\end{figure*}

The different speeds of the algorithms imply an inherent difficulty in comparing the trajectories \textit{point-by-point}. Our analysis will thus rely on defining a particular direction $\hat{v}$ in the space of interactions that we will use to compare the two trajectories and their gradients. Such a direction is defined by the line that connects the initial Hebbian matrix with the point of convergence of the SP,  
\begin{equation}
    \label{eq:v}
    \hat{v} = \frac{\vec{J}_{SP}^{(t_{max})}/\sigma_{SP}^{ (t_{max})} - \vec{J}^{(0)}/\sigma^{(0)}}{| \vec{J}_{SP}^{(t_{max})}/\sigma_{SP}^{ (t_{max})} - \vec{J}^{(0)}/\sigma^{(0)}|} \ .
\end{equation}
We can now define two time-dependent observables that can help us in the analysis of the trajectories:
\begin{equation}
    \label{eq:h}
    q_{v}(t) =  \frac{\vec{J}^{(t)}/\sigma^{(t)} - \vec{J}^{(0)}/\sigma^{(0)}}{|\vec{J}^{(t_{max})}/\sigma^{(t_{max})} - \vec{J}^{(0)}/\sigma^{(0)}|} \cdot \hat{v} \ ,
\end{equation}
which is a measure of the angular distance of any point of the trajectory from the line traced by the direction $\hat{v}$ at time $t$, with $q_v(t)\in [0,1]$ $\forall t$. The smaller this quantity is, the more evidently the trajectory is diverging from $\hat{v}$. One can also introduce
\begin{equation}
    \label{eq:q_grad}
    q_{\Delta, v}(t) =  \frac{\vec{J}^{(t+1)}/\sigma^{(t+1)} - \vec{J}^{(t)}/\sigma^{(t)}}{|\vec{J}^{(t+1)}/\sigma^{(t+1)} - \vec{J}^{(t)}/\sigma^{(t)} |} \cdot \hat{v} \ ,
\end{equation}
that is, the projection of the variation of $\vec{J}$ at the step $t$ along the direction $\hat{v}$. The larger this quantity is, the more aligned to $\hat{v}$ the trajectory is.

Fig.~\ref{fig:gradients}(left) represents the values of $q_v$ during the same trajectory that is depicted in 
fig.~\ref{fig:trajectories}(left). 
One can see that $q_v(0)$ is small for both the HU and the SP. This means  that they both start in the wrong direction: this is particularly reasonable for the HU, which involves a random picking of the initialization state. However, while the SP rapidly reaches $q_v = 0$ because of its high initial acceleration at all the considered values of $\lambda$, in the HU algorithm $q_v$ is decreasing at lower rate. An initial overshooting of the SP at high values of $\lambda$ is signaled by an anomalously high value of $q_v$ at the second step of the process.

The directions followed by the two algorithms with respect to $\hat{v}$ are shown in Fig.~\ref{fig:gradients}(center). The SP at $\lambda = 10^{-4}$ has a peak in $q_{\Delta,v}$ in the first part of the trajectory, signaling a high degree of alignment between the gradient and $\hat{v}$. Later on, the SP rapidly converges towards the condensed region, where gradients lose their polarization with $\hat{v}$, but the convergence point has already been reached. When higher values of $\lambda$ are used, no relevant polarization is measured. The HU shows a similar behavior: the trajectory starts along a direction that is barely aligned with $\hat{v}$ but a consistently high degree of alignment is obtained after more or less half of the iterations. Eventually, the HU also converges towards the final state losing the alignment with $\hat{v}$. Fig.~\ref{fig:gradients}(right) displays the direction of the variation as a function of the distance from $\hat{v}$. Three different behaviors of the trajectory can be thus recognized for both algorithms. First the trajectory moves away from $\hat{v}$ after a bad start, in a second phase it aligns to $\hat{v}$, and in a third phase the matrix plunges towards the convergence state.

\section{\label{sec:5}Geometric Interpretation of Unlearning}

\begin{figure}
\centering
\includegraphics[width=.5\textwidth]{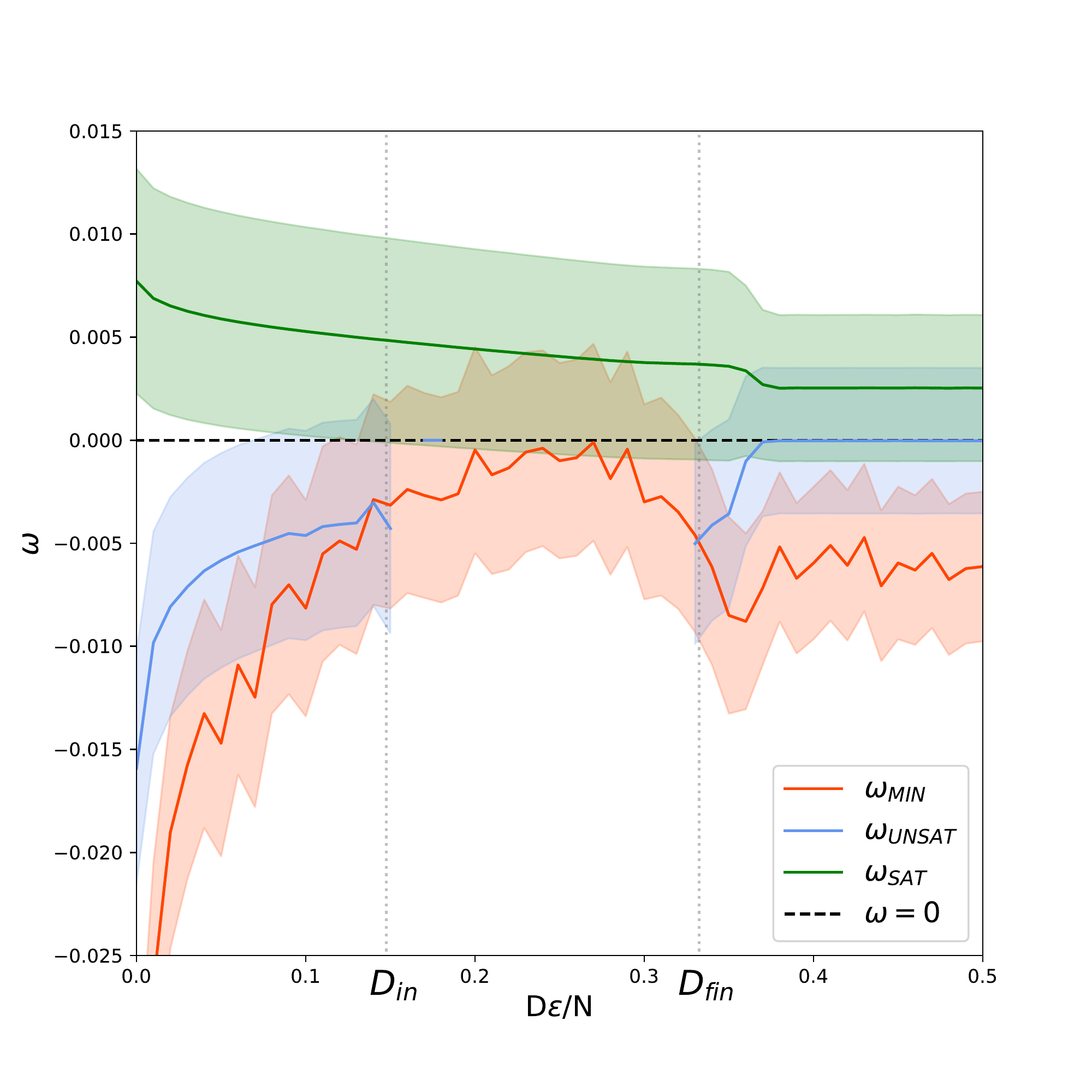}
    \caption{Mean values assumed by the perceptron overlap $\omega$ for SAT, UNSAT and $min$ patterns during the Hebbian unlearning process. Measurements are performed at $N = 800$, $\alpha = 0.3$, $\epsilon = 10^{-2}$ over $200$ spurious states and several realizations of the disorder. Error bars are indicated by the shaded region.}
\label{fig:percept_overlap}           
\end{figure}

In order to provide an argument that might explain the similarities between the HU and SP algorithms, we rewrite the rule in Eq.~(\ref{eq:unlearning_algorithm}) in a vectorial fashion,
\begin{equation}
    \label{eq:vec_unlearning}
    \vec{J_i}^{(D+1)} = \vec{J_i}^{(D)} - \frac{\epsilon}{N}\vec{\eta_i}^*\ ,
\end{equation}
where $\vec{J}_i$ is the vector of the elements contained in the $i^{th}$ row of the connectivity matrix and we call $\vec{\eta_i}^*$ a \textit{glassy pattern}, defined in analogy with the memory patterns $\vec{\eta_i}^{\mu}$ as
\begin{equation}
    \label{eq:glassy_pattern}
    \vec{\eta_i}^* = \sigma_i^*\vec{\sigma}^*\ ,
\end{equation}
being $\vec{\sigma}^*$ the spurious state to which the unlearning algorithm converges.
We also introduce the perceptron overlap
\begin{equation}
    \label{eq:percept_overlap}
    \omega_i^{\mu} = \langle \overline{\frac{1}{N}\vec{\eta_i}^* \cdot \vec{\eta_i}^{\mu}} \rangle\ ,
\end{equation}
where the overbar indicates an average over the spurious states in a given realization of the disorder. Each pair $(i,\mu)$ is thus related to a given constraint of the associated optimization problem, with $\Delta_i^{\mu} \geq 0$ for SAT constraints, and $\Delta_i^{\mu} < 0$ for UNSAT ones. Fig.~\ref{fig:percept_overlap} shows $\omega_i^{\mu}$ at $\alpha = 0.3$ and $N = 800$ for the three types of constraints: SAT, UNSAT and \textit{minimally satisfied}, i.e. the SAT constraints with the lowest measured stability. The fact that the perceptron overlap is negative for both UNSAT and \textit{min} constraints, but positive for SAT constraints, suggests that the distribution of the $\vec{\eta}_i^{\mu}$ looks anisotropic from the reference frame of the glassy patterns. This is certainly induced by the fact that glassy patterns $\vec{\eta_i}^*$ are SAT by definition, so they are more likely to be contained in the same half of hyperspace, defined by the orthogonal plane to $\vec{J}_i$, that contains SAT memory patterns. 

Moreover, since there is a minus sign on r.h.s. of Eq.~(\ref{eq:vec_unlearning}), HU is performing the same geometric transformation of the perceptron in order to align the $\vec{J}_i$ vectors to the memory patterns $\vec{\eta_i}^{\mu}$. By only exploiting the landscape of the spurious states of the Hopfield model out of the retrieval regime, the HU algorithm manages to accomplish this task in an optimal way. We suppose that the small, but yet non null, overlap of spurious states with the memories is an important feature to ensure the maximization of the size of the basins of attraction. 

\section{\label{sec:6}Conclusions}

Both the first part of this paper and Refs.~\cite{VH90,VH92,vanhemmenHebbianLearningIts1998} analyze the effect of Hebbian unlearning on the Hopfield model, giving a measure of the optimal amount of iterations that maximizes the performances of the network in terms of the size of the basins of attraction and memory retrieval. In particular, our present results focus on the case where asynchronous dynamics is performed, the activity of the memories is homogeneous on the network, autapses and dilution are absent in the graph.
By basing our analysis on the study of the minimum stability reached by the memories we were able to give new insights into the classification capabilities of the network. We have defined three relevant amounts of iterations $D_{in}$, $D_{top}$ and $D_{fin}$ and we related them to the average radius of the basins of attraction. We found that the optimal amount of steps, i.e. the one where all memories are perfectly recalled and basins are maximal in size, is $D_{in}$, at variance with~\cite{Horas} where the optimal state of the network was assumed to coincide with the point where the highest minimum stability was reached, in accordance with~\cite{krauth}. This relevant quantity scales as the optimal number of iterations measured in \cite{VH90,VH92,vanhemmenHebbianLearningIts1998} at leading order in the system size $N$, but it appears to be smaller by a correction $O(\alpha^{1/2})$. Results obtained from this kind of analysis are confirmed by the estimation of the critical capacity for the HU, which is perfectly consistent with previous results~\cite{VH90,VH92,vanhemmenHebbianLearningIts1998}. 

Moreover, we have shown that HU performs consistently with the SP near the maximal stability $k$. This result suggests the HU to be an optimal unsupervised algorithm in terms of generalization of the network: from a Hebbian perspective large basins of attraction imply the capability of the model to associate more exotic stimuli to known memories, i.e. a higher recognition power with respect to new inputs. According to Fig.~\ref{fig:mf_mi}, a SP at $k = 0$ has no generalization capabilities, meaning that it falls into an over-fitted regime, where memories are recognized only if the initial state of the dynamics coincides with the memory itself. An increasing value of $k$ is related to an increase in generalization. HU is able to reach the highest degree of generalization in an unsupervised fashion. 

In the second part of the paper the trajectories of the matrix $J$ in the space of interactions have been considered. The final states of the algorithms maximize the overlap between matrices, suggesting that the algorithms converges to the same regions of the interaction space. In the middle part of the trajectory the two algorithms also explore nearby regions, suggesting that they both follow well overlapping gradients in the space of the couplings. 

The pioneering investigation by Van Hemmen and collaborators~\cite{VH90,VH92,vanhemmenHebbianLearningIts1998} concluded that HU was able to remove correlations among the stored memories, turnining memories into fixed points of the dynamics and enlarging the basins of attraction. Regarding this point we remark the brilliant idea contained in Ref.~\cite{VH94}, where a slightly modified version of the unlearning procedure was proved to partially align with the rule proposed in Ref.~\cite{PS}. Nevertheless, since the network of Ref.~\cite{PS} tends to the pseudo-inverse connectivity matrix~\cite{personnaz}, poorer generalization performances, i.e. smaller basins of attraction, should be expected \cite{kanter}.
Our work suggests an alternative geometric interpretation of Hebbian unlearning in its original version. In Sec.~\ref{sec:5} we have shown that the geometric transformation accomplished by the unlearning rule is very similar to the one performed by a linear perceptron, in particular, a perceptron feeded with noisy versions of the memories. When the algorithm probes spurious states having vanishing in $N$, yet non-null, overlap with the stored memories, weights (i.e. rows of the connectivity matrix) become more aligned with the unstable patterns, favoring their correct classification. Hence, while the geometric transformation itself permits to reach the perfect retrieval of the memories, the noise added in the process implies maximally wide basins of attraction. We remark that this effect is a consequence of the attractors landscape in a Hopfield-like network alone, being the procedure completely unsupervised.
Hence a very close analogy between the effects of these two formally different rules has emerged. 

We conclude by some highly speculative considerations on possible implications of our work.
Our results shed light on a substantial mechanism that might help to understand real neuro-physiological processes lying behind synaptic development in the brain~\cite{dragoi} and dream-sleep in mammals~\cite{sleep}. In the context of dream-sleep, it is important to remind that Hebbian unlearning has been introduced simultaneously with another remarkable contribution by Crick and Mitchinson~\cite{dreams}. Their paper conjectured a sort of \textit{reverse learning} procedure which assigned, for the first time, a biological function to dreams, overcoming their description as mere epiphenomena of neural activity. Such a procedure strongly resembles Hopfield's unlearning. Kinouchi and Kinouchi~\cite{KK} provided some biological examples that might encourage to investigate this type of synaptic transformation, even though a clear evidence of its existence has not been shown yet. Furthermore, a recently published review by Hoel~\cite{hoel} corroborates the evolutionary significance of dreams in terms of the generalization performances of neural networks. Dreams, due to their hallucinoid contents, are responsible for noise injection in the learning procedure, in analogy with dropout~\cite{drop} techniques used in machine learning: a dreaming neural network is thus able to generalize better, avoiding over-fitting. The importance of noise addition in learning is also suggested by other recent studies that try to increase generalization in deep neural networks by taking inspiration from biology~\cite{tadros,audio}. According to these works, a local Hebbian-like action on synapses can ensure decorrelation of the stored memories, and thus avoid confusion. We do believe that what we found in the Hopfield model is coherent with such a picture: Hebbian unlearning is not a form of \textit{reverse} learning, as repeatedly stated in the past literature, but it is rather responsible of the learning of noisy versions of the memories, which help to minimize over-fitting. 

One possible development of this research might deal with memories presenting strong structural correlations such as images. In this case a linear regression operation, as the one performed by linear perceptrons, may not be sufficient to ensure classification. It has been shown that Hebbian unlearning works well even with digits~\cite{VH92}. We suggest that such correlations, being encoded in the quenched disorder of the system, and thus in the glassy landscape of attractors, might drift the learning path right to the optimal region of the space of interactions. Another direction for future work would be the verification of the results on other types of models, such as more biologically reliable Hopfield-like networks~\cite{treves}, random neural networks \cite{hwang19,hwang20}, or continuous attractor neural networks~\cite{battista}. This last class of systems, which aim at describing the functioning of the spatial memory encoded in the hippocampal synapses, might open the way to experiments and inferential analyses of real data, allowing a proper research of physiologic unlearning-like mechanisms in the brain. 

Finally, these ideas could possibly find application and analogies in the training of physical systems, such as meta-materials or allosteric networks, see e.g.~\cite{nagel1,nagel2}.

\begin{acknowledgments}
We thank Dario Lippi for his important contribution during the first stage of this work.
\end{acknowledgments}



\nocite{*}

\bibliography{biblio}

\end{document}